\documentclass[preprint,preprintnumbers,amsmath,amssymb]{revtex4}

\def\be{\begin{equation}}
\def\ee{\end{equation}}
\def\beq{\begin{eqnarray}}
\def\eeq{\end{eqnarray}}

\def\bay{\begin{array}}
\def\eay{\end{array}}

\begin{document}
\title{Towards a Universal Theory of
Relativity}

\author{Sanjay M. Wagh}
\affiliation{Central India Research Institute, \\ Post Box 606,
Laxminagar, Nagpur 440 022, India\\
E-mail:cirinag\underline{\phantom{n}}ngp@sancharnet.in}

\date{March 5, 2005 \\ \phantom{m} \vspace{.5in}}
\begin{abstract}
We discuss here the significance of the generalization of the
newtonian concept of force by that of a transformation of a
certain Standard Borel Space of cardinality $\mathbf{c}$ of the
continuum as the ``cause'' behind motions of material bodies that
are representable as Borel measurable subsets of this space. This
generalization forms the basis for a Universal Theory of
Relativity in which, importantly, the fundamental physical
constants can only arise from mutual relationships of the
so-defined physical bodies. This Universal Relativity also has the
potential to explain the quantum nature of the physical world.
\vspace{.5in} \\
\noindent \centerline{(Essay for Gravity Research Foundation
Competition - 2005)}
\end{abstract}

\maketitle \newpage

Galileo's concept of inertia is the foremost of the concepts
behind Newton's theory which postulates that every material body
has inertia for motion.

Newton's theory represents physical bodies as points of the space
$\mathbb{R}^3$ with associated inertias and Newton's Three Laws
describe their motions. Then, the Euclidean distance is the {\em
physical distance\/} and $\mathbb{R}^3$ is the {\em physical
space\/} for this theory.

A newtonian observer can {\em observe\/} the {\em entire
trajectory\/} of a material body without disturbing it. A {\em
physical clock\/} is a material body undergoing periodic motion.
{\em Independently\/} of other material points and the
coordination of the metrically flat space $\mathbb{R}^3$, an
``exact measurement'' of the state of the clock {\em yields\/} the
{\em physical time}. An observer can check the position of a
chosen material point against the {\em simultaneous\/} state of a
physical clock body. The motion of a material body, including that
of the clock body, does not produce any change in the Euclidean
space $\mathbb{R}^3$ or in its coordination.

Now, any difference in coordinates of $\mathbb{R}^3$ is a
``measuring stick'' for physical measurements of distances. Each
observer has associated measuring sticks and clocks in Newton's
theory. Then, the {\em entire\/} system of measuring sticks and
clocks is carried with that observer when in motion relative to
another observer.

Clearly, two material sticks {\em cannot occupy} the same place.
But, in Newton's theory, a measuring stick of one observer does
not collide with that of another observer in motion even when both
these sticks arrive at the same place. Unacceptably, measuring
sticks just pass through each other without colliding on their
first contact.

But, the same situation does not arise for material points which
collide on contact. Then, in Newton's theory, measuring sticks and
clocks are treated {\em separately\/} than all the other physical
objects. This is very disturbing and unsatisfactory.

Next, the force, as a cause of motion, is another pivotal concept
of Newton's theory. Only a material point can be the source of
force, and, consequently, Newton's is an {\em action at a
distance\/} framework.

A {\em total force\/} acting on a body then provides the means of
establishing its path on the basis of Newton's Second Law of
Motion. Notably, without the {\em Law specifying Force}, the {\em
Law of Motion\/} is {\em empty\/} of contents in Newton's theory.

From our day-to-day observations, we notice that various objects
{\em gravitate\/} towards the Earth, {\em ie}, the distance
between them decreases with time as observed.

Newton ``explained'' {\em gravitation\/} by {\em postulating\/} an
{\em attractive force of gravity\/} that is proportional to {\em
inertias\/} of two material points varying inversely with the
square of distance separating them. It is a {\em universal\/}
force, as inertia characterizes bodies. Newton's theory
``explains'' observations when his Third Law of Motion is also
assumed.

In Newton's theory, a material body in fact has two {\em
independent\/} attributes: the first, its {\em inertia}, and the
second, its {\em gravitational mass}. However, various
observations, since Galileo's times, then indicate \cite{0411052}
that the inertia and the gravitational mass of a material body are
equal to a high degree of accuracy. However, this equality as well
as the inverse-square dependence of the gravitational force on the
distance separating two bodies become assumptions of Newton's
theory.

Now, every object does not fall to the Earth. This is then
explainable by postulating forces opposing attractive gravity.
However, in Newton's theory, {\em every\/} (basic) {\em force\/}
is a {\em postulate\/} needing an \underline{assumed}
\underline{source} p\underline{ro}p\underline{ert}y attributable
to physical bodies. Notably, every action-at-a-distance force has
this characteristic always. Then, action-at-a-distance theories
cannot ``explain'' the origins of {\em assumed\/} source
properties.

Such theoretical reasons as well as many well known experiments
demand a new theory that must, fundamentally, abandon some
newtonian concepts.

Now, force of gravity needs a source. Although conceptually
different, this source-mass ``equals'' inertia in value for every
body. Hence, inertia is a more general concept than the force,
with only the latter then coming under scrutiny for abandonment.

For a physical description of the phenomena displayed by Light,
Einstein assumed the {\em special\/} principle of relativity,
which is essentially the {\em same\/} as the newtonian principle
of relativity. Assuming also the constancy of the speed of Light
for all inertial observers, he then developed the Special Theory
of Relativity. This theory is an {\em extension\/} only of the
newtonian laws to incorporate the laws of motion for Light
\cite{100-yrs}.

But, Special Relativity also suffers from problems of treating the
measuring sticks and clocks separately from all other objects.
Following Mach, Einstein then extended \cite{ein-dover} its basis
to the g\underline{eneral} p\underline{rinci}p\underline{le}
\underline{of} \underline{relativit}y that: {\em The laws of
physics must then be such that they apply to systems of reference
in any kind of motion}.

{\em \underline{Clearl}y, \underline{the} \underline{same}
\underline{laws} \underline{o}f p\underline{h}y\underline{sics}
\underline{must} \underline{also} \underline{be} \underline{such}
\underline{as} \underline{to} \underline{allow} \underline{the}
\underline{re}f\underline{erence} f\underline{rames}
\underline{to} \underline{be} \underline{a}ff\underline{ected}
\underline{b}y \underline{motions} \underline{o}f
\underline{other} \underline{material} \underline{bodies}}.

Einstein connected the general principle of relativity with the
situation that a possible {\em uniform\/} gravitation imparts the
same acceleration to all bodies and arrived at the well known
(Einstein's) equivalence principle which reduces the equality of
inertia and gravitational mass of a body to a redundancy.
Gravitational mass is, clearly, {\em irrelevant\/} when the
concept of force is abandoned. Only the concept of inertia of a
material body remains relevant to its motion perceived by an
observer.

The general principle of relativity deals only with {\em
observable\/} concepts and stands even when the concept of force
is abandoned. For gravity, it rests on the possibility of uniform
gravity imparting the same (observable) acceleration to all the
bodies.

Einstein, while developing these ideas, wrote \cite{ein-dover}
that ``... in pursuing the general theory of relativity we shall
be led to a theory of gravitation, ... .''

True this. But, the general principle of relativity can be reached
from more than one vantage issues. Each such issue can then
indicate only that some physical phenomenon related to that issue
is {\em consistent\/} with this principle of relativity. For
example, the equivalence principle establishes the {\em
consistency\/} of only the phenomenon of gravitation with the
general principle of relativity.

Clearly, the abandonment of the concept of force applies to
``every (fundamental) force'' that needs to be postulated to be
acting between the chosen material bodies to ``explain''
observations, using Newton's theory or any other theory.

For example, the mathematical procedure by which we replace the
notion of, say, Newton's gravitational \underline{force} {\em
cannot\/} be {\em different\/} than the one adopted, say, for
replacing the notion of Coulomb's electrostatic \underline{force}.

Therefore, the {\em conceptual framework\/} and, hence, also the
{\em mathematical formalism}, which ``replaces'' the concept of
force {\em will have to be applicable to every (fundamental)
force\/} that Newton's theory or any other theory has to postulate
to ``successfully'' explain the observed phenomena.
\underline{This} \underline{is}
\underline{conce}p\underline{uall}y \underline{mandator}y.

Then, a physical theory based on the general principle of
relativity, call it the \underline{Universal} \underline{Theor}y
\underline{of} \underline{Relativit}y to differentiate it from
Einstein's General Relativity that is only a Theory of
Gravitation, will necessarily be a {\em Theory of Everything}.

Such a theory can, for example, ``explain'' the phenomenon of
gravitation by demonstrating that the decrement of distance
between material bodies is independent of their material contents
and physical state {\em in situations for which the corresponding
total force on bodies is that given by the newtonian law of
gravity}.

Now, the mathematical framework of the Universal Theory of
Relativity \underline{cannot} be based on the ``orthodox rules''
of the Quantum Theory.

Quantum Theory provides us only the means to determine
(Schr\"{o}dinger's) $\Psi$-function and thereby obtain the
probability of a physical event involving physical object(s) when
we \underline{s}p\underline{ecif}y, \underline{b}y
\underline{hand}, either the lagrangian or the hamiltonian, {\em
ie}, certain physical properties, for physical bodies under
consideration. Then, ``origins'' of, {\em need to be assumed},
``intrinsic physical properties'' of bodies \underline{cannot} be
explainable by adopting mathematical rules of the Quantum Theory
for the Universal Theory of Relativity.

Now, the concept of force is, in a definite mathematical sense
\cite{dyn-sys}, \underline{e}q\underline{uivalent} to that of
certain transformations of the point of the (Euclidean) space
$\mathbb{R}^3$ in Newton's theory. This observation is then
suggestive that mathematical transformations of points of some
suitable (underlying) space can, quite generally as well as
naturally, ``\underline{re}p\underline{lace}'' the newtonian
concept of force as a cause of motion.

Evidently, the physical laws obtained by using this generalization
will be applicable to {\em every reference frame}, and, hence,
this mathematical formalism will be in conformity with the general
principle of relativity.

Consequently, confirmed results of Newton's theory as well as
those of the Special Theory of Relativity will, evidently, be
\underline{obtainable} in the Universal Theory of Relativity by
treating the involved (newtonian) forces as corresponding
transformations of the suitable underlying space of the Universal
Theory of Relativity.

But, it must follow from the mathematical formalism of the
Universal Relativity that the ``inertia'' can also be
``naturally'' considered as the ``source'' in the mathematical
quantity that can be the newtonian gravitational force.

Similarly, the quantity that, in the Universal Theory of
Relativity, replaces the electrostatic charge must also naturally
appear as the ``source'' in the mathematical quantity that can be
considered to be Coulomb's electrostatic force.

Thus, we arrive at the important issue of selecting an appropriate
underlying space and associated mathematical formalism for the
Universal Theory  of Relativity whose certain characteristics we
have been considering above.

Then, let the physical world, excluding time, be based on a
3-dimensional space that we shall denote by $\mathfrak{S}$. We
call $\mathfrak{S}$ the p\underline{h}y\underline{sical}
\underline{s}p\underline{ace} \underline{underl}y\underline{in}g
\underline{Universal} \underline{Relativit}y. We explore below
mathematical properties \cite{measure-theory} permissible for it.

A separable, completely metrizable topological space $X$ is a
\underline{Polish} \underline{s}p\underline{ace}. A
\underline{measurable} \underline{s}p\underline{ace} is a pair
$(X,\mathcal{A})$ with $\mathcal{A}$ being a $\sigma$-algebra of
the subsets of the set $X$. Members of the $\sigma$-algebra
$\mathcal{A}$ are called \underline{measurable} \underline{sets}.
If $\mu$ is a \underline{measure} on $\mathcal{A}$, we call the
triplet $(X,\mathcal{A},\mu)$ a \underline{measure}
\underline{s}p\underline{ace}. Notably, concepts of measure theory
\cite{measure-theory} hold (mod 0), {\em ie}, when sets of measure
zero are discarded from its considerations.

Smallest $\sigma$-algebra containing the topology $\mathcal{T}$ on
set $X$ of a measurable space $(X,\mathcal{A})$ is a
\underline{Borel}
\underline{$\sigma$}-\underline{al}g\underline{ebra} and we write
$\mathcal{B}_X$ for it. Sets in $\mathcal{B}_X$ are
\underline{Borel} \underline{sets} \underline{in} \underline{$X$}.

\underline{Standard} \underline{Borel}
\underline{S}p\underline{ace} is {\em isomorphic\/} to a Borel
subset of a Polish space. Borel set of a Standard Borel Space is
\underline{Standard} and measure on it, a \underline{Borel}
\underline{Measure}.

Now, we \underline{assume} \underline{that} \underline{some}
\underline{suitable} \underline{continuum} \underline{underlies}
\underline{the} \underline{descri}p\underline{tion} \underline{of}
\underline{the} p\underline{h}y\underline{sical}
\underline{world}, {\em ie}, the cardinality of the physical space
$\mathfrak{S}$ is $\mathbf{c}$ and that the space $\mathfrak{S}$
is a \underline{Lebes}g\underline{ue}
\underline{s}p\underline{ace}, {\em ie}, complete (mod 0) relative
to (one of) its basis. Then, {\em physical objects\/} are {\em
Borel subsets\/} of $\mathfrak{S}$ with {\em Borel measures\/}
being their {\em physical properties}.

A p\underline{artition} (\underline{mod} \underline{0}) of measure
space $(X,\mathcal{A}, \mu)$ is any family $\Upsilon =\{C_i:i \in
I\}$ of nonempty disjoint subsets of $X$ such that
$\bigcup_iC_i=X$ (mod 0). The sets $A\in \mathcal{A}$ which are
the unions of the members of $\Upsilon$ are called
\underline{$\Upsilon$}-\underline{sets}.

Now, let us call every member of a \underline{measurable}
p\underline{artition} (\underline{mod} \underline{0}) $\Upsilon$
of the p\underline{h}y\underline{sical} \underline{measure}
\underline{s}p\underline{ace} $(\mathfrak{S},
\mathcal{B}_{\mathfrak{S}}, \mu)$ as a \underline{basic}
p\underline{h}y\underline{sical} \underline{ob}j\underline{ect}.
Hence, any $\Upsilon$-set, also a standard Borel set in
$\mathfrak{S}$, is a \underline{com}p\underline{ound}
p\underline{h}y\underline{sical} \underline{ob}j\underline{ect}.

Then, a transformation of the space $(\mathfrak{S},
\mathcal{B}_{\mathfrak{S}}, \mu)$ can be performed which does not
affect some physical object, some $\Upsilon$-set, whose
``location'' is being determined, but ``moves'' only the measuring
stick, another $\Upsilon$-set, in the manner desired by the
observer for the involved measurement. It is a measure preserving
transformation of $(\mathfrak{S}, \mathcal{B}_{\mathfrak{S}},
\mu)$. Then, if a physical body were representable as ``exactly
localizable material point'' in the framework of Universal
Relativity, {\em ie}, a $\Upsilon$-set ``represented'' by a
singleton subset of $\mathfrak{S}$, we could determine its exact
location on moving measuring stick by its side without affecting
the location of that body since a transformation of the space
$(\mathfrak{S}, \mathcal{B}_{\mathfrak{S}}, \mu)$ that achieves
this, including Light to ``see'' the process, is permissible.

[Newton's theory represents a physical body as an exactly
localizable material point. Then, we can determine the exact
location of this material point because a transformation of
$\mathbb{R}^3$, a force, not affecting the material point but
``moving'' only the measuring stick in the desired manner is
permissible in this theory.]

This would, however, violate Heisenberg's celebrated indeterminacy
relations (\cite{heisenberg-bohr}. See, also, Bohr N in
\cite{schlipp}.). Then, assuming the ``correctness'' of
indeterminacy relations, we readily infer that, {\em within
Universal Relativity, it must be impossible to hypothesize an
exactly localizable material point to represent a physical
object}.

Now, a measure averaged over any basic or compound physical
object, the average being a property of each of its points,
provides \cite{issues, field-program, indeterminacy, heuristic,
jaipur, results} the non-singular notion of a point object with
the physical characteristics. The ``location'' of the point-object
so defined is {\em intrinsically indeterminate\/} within the
corresponding $\Upsilon$-set.

In Universal Relativity, the action of a (Borel) transformation of
the physical space $\mathfrak{S}$ as a ``cause'' of the motion and
the {\em intrinsically indeterminate\/} location of a material
point are then the keys to quantum aspects of matter.

The {\em physical distance\/} is then an appropriate {\em
mathematical distance\/} between measurable sets. ``Kinematical''
quantities such as ``velocity'' and ``acceleration'' involve
change in the so-defined physical distance under the action of the
Borel transformation of the physical space $\mathfrak{S}$. Various
physical phenomena can then arise from actions of transformations
of $(\mathfrak{S}, \mathcal{B}_{\mathfrak{S}}, \mu)$ on its
measurable sets and the measures defined on them. This is then the
framework of the theory of dynamical systems \cite{dyn-sys}.

Notably, there ``do not occur'' any ``physical constants'' to be
``specified by hand'' in this above framework. But, {\em all\/}
the physical constants can arise in this framework only from
``mutual relationships'' of involved physical objects.

For example, in Universal Relativity, the phenomenon of
gravitation involves the action of transformation $T$ for which
the ``acceleration'' of one measurable set relative to another
reference measurable set is {\em independent\/} of the ``measure''
defined on that set, but is proportional to the measure defined on
the reference set, both measure classes being invariant under $T$.
Newton's gravitational constant $G$ then ``arises'' when
``acceleration'' is expressed as the ``inverse-square'' of the
physical distance. Clearly, the possibility of
\underline{theoreticall}y \underline{obtainin}g the ``value'' of
$G$ arises in this manner.

In Universal Relativity, a physical constant can only ``arise''
from mutual relationships of measurable sets and effects of
measure preserving transformations of the space $\mathfrak{S}$ on
them. (Fundamentally, this is also how we determine these
constants experimentally.) Clearly, the ``values'' of such
physical constants \underline{cannot} be changed and this
situation is, precisely, as per Einstein's related theorem
\cite{schlipp} (p. 63).

Notably, nowhere in Universal Relativity, in its explanations of
physical phenomena, do we require the ``intervention'' by any
``observer,'' conscious or not. {\em Newton's theory also had the
same role for an observer}. Then, a transformation of the space
$\mathfrak{S}$ is a {\em unique evolution\/} of its points and,
hence, it represents a {\em unique evolution of a physical system
``fixed'' deterministically by the initial conditions}.
Consequently, Universal Relativity provides us ``the complete
description of any individual real situation as it supposedly
exists irrespective of any act of observation or substantiation''
\cite{schlipp}.

Lastly, the \underline{uni}q\underline{ue}
\underline{identif}y\underline{in}g \underline{characteristic} of
the physical space $\mathfrak{S}$ is then provided by the
following key physical situation: {\em physical matter can be
assembled (and reassembled) in any arbitrary manner at any
location in the Universe}. But, this is equivalent to changing
{\em continuously\/} measurable partitions and Borel measures of
the physical measure space $\mathfrak{S}$. Perhaps, for this, the
continuum $\mathfrak{S}$ needs to admit three, linearly
independent, homothetic Killing vectors which {\em uniquely
determine\/} it.

Granted the above, mathematical foundations of our fundamental
understanding of the physical world then rest on theories of
measures and dynamical systems.

\acknowledgments I am indebted to S G Dani, M G Nadkarni, R V
Saraykar and V M Wagh for drawing my attention to literature on
theories of measures \& dynamical systems and, to N Krishnan as
well, for many helpful, clarifying and encouraging discussions.

\end{document}